\documentclass[aps,prd,12pt,preprint,tightenlines,superscriptaddress,
amsfonts,amssymb,amsmath,byrevtex,showpacs,nofootinbib]{revtex4}
\usepackage{graphics}
\usepackage{epsfig,subfigure}
\begin{document}
\newcommand{\dR}{\mathbb R}
\newcommand{\dC}{\mathbb C}
\newcommand{\dS}{\mathbb S}
\newcommand{\dZ}{\mathbb Z}
\newcommand{\id}{\mathbb I}
\newcommand{\dM}{\mathbb M}
\newcommand{\dH}{\mathbb H}
\newcommand{\tm}{\tilde{\mu}}
\newcommand{\tn}{\tilde{\nu}}

\title{Excited states of a string in a time dependent orbifold}

\author{Przemys{\l}aw Ma{\l}kiewicz$^\dag$ and W{\l}odzimierz Piechocki$^\ddag$
\\ Department of Theoretical Physics\\Institute for Nuclear Studies,
\\ Ho\.{z}a 69, 00-681 Warszawa, Poland;
\\ $^\dag$pmalk@fuw.edu.pl, $^\ddag$piech@fuw.edu.pl}

\date{\today}

\begin{abstract}
We present analytical results on the propagation of a classical
string in non-zero modes through the singularity of the
compactified Milne space. We restrict our analysis to a string
winding around the compact dimension  of spacetime. The compact
dimension undergoes contraction to a point followed by
re-expansion. We demonstrate that the classical dynamics of the
string in excited states is non-singular in the entire spacetime.
\end{abstract}
 \maketitle

\section{Introduction}

One of the simplest models of the neighborhood of the cosmological
singularity (CS), inspired by string/M theory
\cite{Khoury:2001bz}, is the compactified Milne (CM) space. It has
been used in the cyclic universe scenario
\cite{Khoury:2001bz,Steinhardt:2001vw,Steinhardt:2001st}. Figure 1
shows the two dimensional CM space embedded in three dimensional
Minkowski space. It can be specified by the following isometric
embedding
\begin{equation}\label{emb}
    y^0(t,\theta) = t\sqrt{1+r^2},~~~~y^1(t,\theta) =
    rt\sin(\theta/r),~~~~y^2(t,\theta) = rt\cos(\theta/r),
\end{equation}
where $ (t,\theta)\in \dR^1 \times \dS^1 $ and $ 0<r \in\dR^1 $ is
a constant labelling compactifications . One has
\begin{equation}\label{stoz}
    \frac{r^2}{1+r^2}(y^0)^2 - (y^1)^2- (y^2)^2 =0.
\end{equation}
Eq. (\ref{stoz}) presents two cones with a common vertex at
$\:(y^0,y^1,y^2)= (0,0,0)$. The induced metric on (\ref{stoz})
reads
\begin{equation}\label{line1}
    ds^2 =  - dt^2 +t^2 d\theta^2 .
\end{equation}
Generalization of the 2-dimensional CM space to the $d+1$
dimensional spacetime has the form
\begin{equation}\label{line2}
ds^2 = -dt^2  +  t^2 d\theta^2 + \delta_{kl}~dx^k dx^l ,
\end{equation}
where $t,x^k \in \mathbb{R}^1,~\theta\in \mathbb{S}^1~(k=
2,\ldots, d)$.

One term in the metric (\ref{line2}) disappears/appears at $t=0$,
thus the CM space may be used to model the big-crunch/big-bang
type singularity. Orbifolding  $\dS^1$ to the segment gives a
model of spacetime in the form of two orbifold planes which
collide and re-emerge at $t=0$. Such a model of spacetime was used
in \cite{Khoury:2001bz,Steinhardt:2001vw,Steinhardt:2001st}. Our
results apply to both choices  of topology of the compact
dimension.

The CM space is an orbifold due to the vertex at $t=0$. The
Riemann tensor components equal $0$ for $t\neq 0$. The singularity
at $t=0$ is of removable type: any time-like geodesic with $t<0$
can be extended to some time-like geodesic with $t>0$. However,
the extension cannot be unique due to the Cauchy problem at $t=0$
for the geodesic equation  (the compact dimension shrinks away and
reappears at $t=0$).
\begin{figure}[h]
\includegraphics[width=0.3\textwidth]{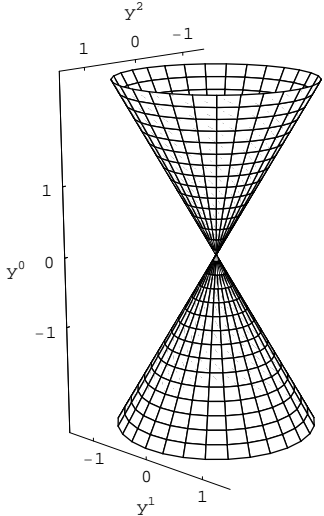}
\caption{Compactified 2d Milne space embedded in 3d Minkowski
space.}
\end{figure}

The CS plays the key role, in the cyclic model of the evolution of
the universe, because it joins  each two consecutive classical
phases. A reasonable model of the CS should allow for the
propagation of an elementary object (particle, string,
membrane,...)  from the pre-singularity to the post-singularity
epoch. If the CS constitutes an insurmountable obstacle for
elementary objects, the cyclic evolution cannot be realized.

We have already applied the above criterion for the propagation of
a test particle \cite{Malkiewicz:2005ii,Malkiewicz:2006wq} and a
test string \cite{Malkiewicz:2006bw}. In what follows (and in
\cite{Malkiewicz:2006bw}) we examine the dynamics of a string in
the so-called winding mode \cite{Pioline:2003bs,Turok:2004gb}. It
is defined to be a state in which the string is winding around the
compact dimension undergoing contraction to a point followed by
re-expansion. In \cite{Malkiewicz:2006bw} we have considered the
classical and quantum dynamics of a {\it zero-mode} string, i.e. a
string in its lowest energy state. Here we present results
concerning the propagation of a classical string in {\it non-zero
modes}, i.e. excited states of a string.

The propagation of a string is described by analytic functions.
Thus, it is non-singular in the entire spacetime including the CS.
The results we have obtained suggest that the CM space is a
promising model of the CS deserving further investigation.

The next section presents the method of finding solutions to the
dynamics in the CM space. First, we define 2d CM space by making
use of 2d Minkowski space. Then, we recall known solutions in the
Minkowski space. Later, we impose the topology and symmetry
conditions, specific for the CM space,  on the solutions in the
Minkowski space. Realization of these conditions leads finally to
the solutions in the CM space. As illustrations, we give two
specific examples of solutions. We conclude in the last section.

\section{Dynamics of a string}

\subsection{Local flatness of the compactified Milne space}

The metric corresponding to the compactified Milne space reads
\begin{equation}\label{met}
g_{\mu\nu}dx^\mu dx^\nu = -dt^2+t^2d\theta^2+d\overline{x}^2 .
\end{equation}
It is locally flat and can be rewritten in the form of the
Minkowski metric, if we make the following change of coordinates
\begin{equation}\label{min}
    x^0 = t\cosh{\theta},~~~~~~x^1 = t\sinh{\theta} .
\end{equation}
Figure 2 illustrates the compactification of the Milne space.
Suppose that $(t,\theta)$ are coordinates of the Milne space,
which parameterize the interior of the light-cone. We compactify
the Milne space by identification of the points $\theta \sim
\theta+\beta $ for some fixed value of $\beta$, where $\theta\in
[0,\beta[$. This way the boundaries of the `grey' region become
identified.

\begin{figure}[h]
\includegraphics[width=0.3\textwidth]{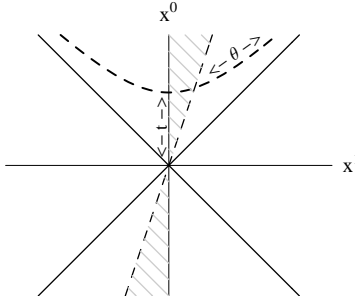}
\caption{The interior of the light cones defines the covering
space of the compactified Milne space.}
\end{figure}

In what follows we use the {\it local flatness} of (\ref{met}) to
solve the dynamics of a string in the CM space.

\subsection{Solutions in the Minkowski space}

The string propagation is expressed in terms of embedding
functions $x^\mu(\tau,\sigma)$, which map the 2d space of the
string coordinates into the $(d+1)$-dimensional CM space. An
action describing a test string in a fixed background spacetime
with metric $g_{\mu \nu}$ may be given by  the Polyakov action
\begin{equation}\label{ac2}
S_{P} = -\frac{1}{2}\mu\int d\tau d\sigma
~\sqrt{-\gamma}~\gamma^{ab}~x_{,a}^\mu x^\nu_{,b}~g_{\mu\nu} ,
\end{equation}
where $(\tau,\sigma)$ are string worldsheet coordinates,  $\mu$ is
a mass per unit length, $\gamma_{ab}$ is the string worldvolume
metric, and $\gamma := det[\gamma_{ab}]$.

Inserting $\sqrt{-\gamma}~\gamma^{ab}:=\eta^{ab}$ (which is a
special choice of gauge on the string's worldsheet)  and
$g_{\mu\nu}:=\eta_{\mu\nu}$ into (\ref{ac2}) gives the Minkowski
space case.  Variation of $S_P$ with respect to $x^{\nu}$ gives
\begin{equation}
\delta S_{P} = \mu\int d\tau d\sigma \;(x^\mu_{,\tau}\delta
x^\nu_{,\tau} - x^\mu_{,\sigma}\delta
x^\nu_{,\sigma})~\eta_{\mu\nu} .
\end{equation}
The condition $~\delta S_{P} = 0~$ leads to
\begin{equation}\label{eom}
\partial_{\tau}^2x^{\mu} - \partial_{\sigma}^2x^{\mu} =0 ,
\end{equation}
plus a boundary term. Hence, the string's propagation in Minkowski
space is described by
\begin{equation}\label{mink}
  x^{\mu}(\tau,\sigma) = x^{\mu}_+(\tau+\sigma)+ x^{\mu}_-(\tau-\sigma),
 \end{equation}
\begin{equation}\label{gauge2}
  \partial_{\tau}x^{\mu}\partial_{\tau}x_{\mu}
  +\partial_{\sigma}x^{\mu}\partial_{\sigma}x_{\mu} =0,~~~~~~
\partial_{\tau}x^{\mu}\partial_{\sigma}x_{\mu}=0 ,
\end{equation}
where $x_\pm^\mu$ are any functions. The equations (\ref{gauge2})
are just gauge constraints. We can make use of these solutions to
construct string solutions in the CM space which wind round the
compact dimension, and so can be expressed in terms of a function
$\overline{x}(t,\theta)$, where $\overline{x}:=(x^2, x^3,
\texttt{\ldots}, x^d)$.

\subsection{Topology condition}

It follows from (\ref{min}) that the range of this mapping has a
non-trivial topology due to the existence of the singular point
$(x^0,x^1) = (0, 0)$. If this point is chosen to correspond to
$\tau = 0$ then we arrive at the following topology condition
\begin{equation}\label{sym}
  x^0 = f(\tau+\sigma)-f(-\tau+\sigma),~~~~~
  x^1 = g(\tau+\sigma)-g(-\tau+\sigma) ,
\end{equation}
where $f$ and $g$ are any functions.

More generally, we can perform such conformal transformation
(i.e.,
$\sigma_{\pm}\rightarrow\widetilde{\sigma_{\pm}}(\sigma_{\pm})$,
where $\sigma_{\pm}=\sigma\pm\tau$) on the solution (\ref{mink})
which leads to $x^0 = f(\tau+\sigma)-f(-\tau+\sigma)$. One can
verify that the solutions for $x^0$ which depend only on a single
variable, either $\sigma_+$ or $\sigma_-$, are excluded. It
follows from (\ref{min}) that we have the implication:
$(x^0=0)\Rightarrow (x^1=0)$. This means that for $\tau=0$ we have
$x^1=0$, which leads to $x^1 = g(\tau+\sigma)-g(-\tau+\sigma)$.

The map $(\tau, \sigma)\longrightarrow ( x^0, x^1)$ is everywhere
invertible except on the curve $(\tau=0,\sigma\in \mathbb{R})$,
because this curve is mapped into the single point $(x^0, x^1)=(0,
0)$. In this way we have defined a map with its range in a
neighborhood of the singularity $x^0 =0$, having its domain within
a global coordinate system on the string's worldsheet. It is
illustrated schematically in  Fig 3.

\begin{figure}[h]
\hspace{-0.25\textwidth}
\begin{minipage}[b]{0.55\textwidth}
\flushleft
\includegraphics[width=0.56\textwidth,height=0.56\textwidth]{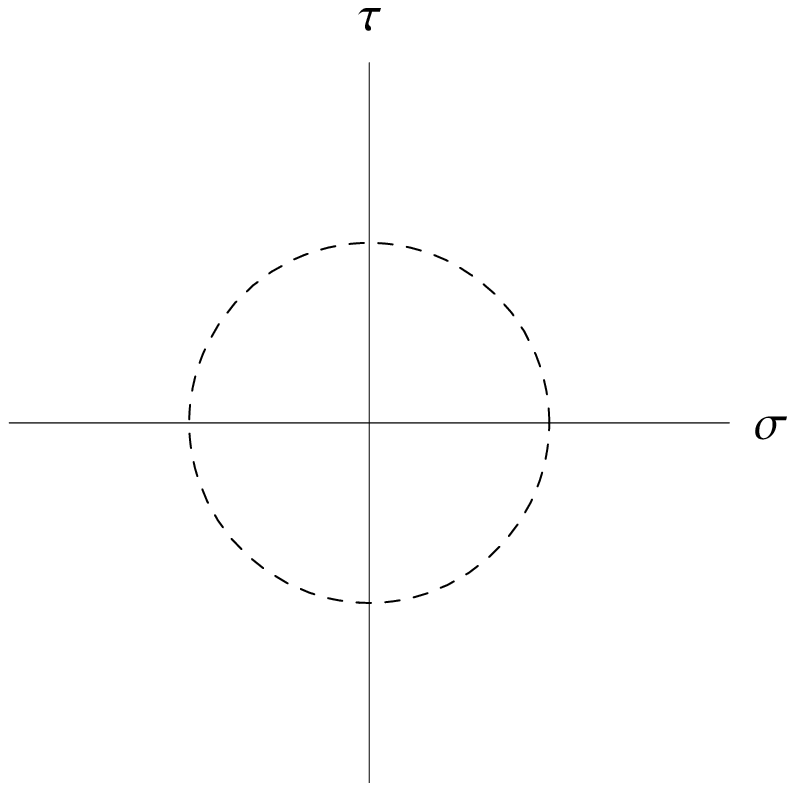}
\nolinebreak
\begin{minipage}[b]{0.32\textwidth}
\includegraphics[width=\textwidth]{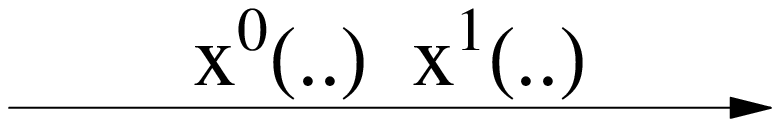}
\vspace{0.6\textwidth}
\end{minipage}
\nolinebreak
\includegraphics[width=0.56\textwidth,height=0.56\textwidth]{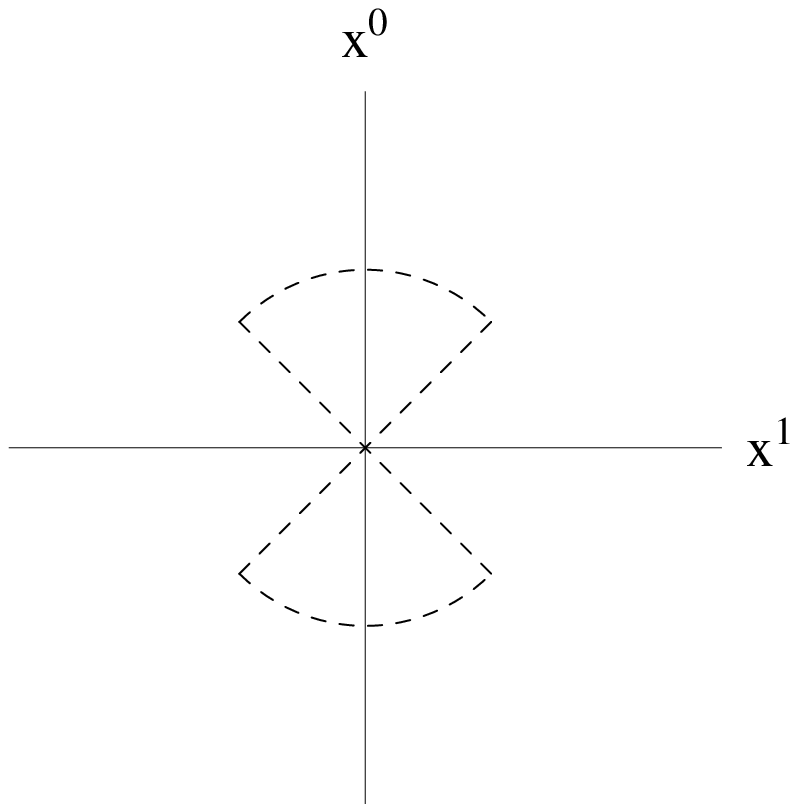}
\end{minipage}
\caption{Singular property of the map $(\tau,
\sigma)\longrightarrow ( x^0, x^1)$. The map is invertible for
$\tau \neq 0$, and non-invertible for $\tau=0$.}
\end{figure}

\subsection{Symmetry condition}

Now, let us impose the symmetry condition  on the remaining $x^k,
(k>1)$ embedding functions. Due to the assumption made earlier,
$x^k$ are functions of $t$ and $\theta$, i.e.
$x^k(\tau,\sigma)=\widetilde{x}^k(t(\tau,\sigma),\theta(\tau,\sigma))$
and are to be periodic in $\theta$. It follows from (\ref{sym})
that
\begin{equation}\label{fun1}
  \theta = \textrm{arctanh}\Big(\frac{g(\sigma_+)-g(-\sigma_-)}
  {f(\sigma_+)-f(-\sigma_-)}\Big)
\end{equation}

\begin{equation}\label{fun2}
t=\textrm{sgn}(\tau)~\sqrt{(f(\sigma_+)-f(-\sigma_-))^2-(g(\sigma_+)
-g(-\sigma_-))^2}
\end{equation}
So the symmetry condition states that
$x^k=x^{k}_+(\sigma_+)+x^{k}_-(\sigma_-)$ is periodic in
$\theta=\textrm{arctanh}(\frac{g(\sigma_+)-g(-\sigma_-)}{f(\sigma_+)-f(-\sigma_-)})$.
In other words, we should determine $x^{k}_+$ and $x^{k}_-$ from
\begin{equation}\label{deter}
  x^{k}_+(\sigma_+)+x^{k}_-(\sigma_-) = \sum_n a_n^k(t)\exp\big
(\imath\frac{2\pi
  n}{\beta}\theta\big),
\end{equation}
where $a_n^k$ are functions of $t$ whose exact form we will
discover below.  It may seem to be impossible to satisfy these
conditions. One obstacle is due to the fact that on the left-hand
side we have a sum of functions of a single variable, while on the
right-hand side there is a sum of functions which depend in a
rather complicated way on both variables. However, we can compare
both sides of (\ref{deter}) at a line. In this way one can rule
out one of the variables and compare functions dependent on just a
single variable. The procedure rests upon the fact that the
dynamics is governed by a second order differential equation
(\ref{eom}), and thus it is sufficient to satisfy the symmetry
condition by specifying $x^k$, $\partial_t x^k$ on a single
Cauchy's line. We choose it to be the singularity, i.e. the line
$\sigma_+ = -\sigma_-$, or equivalently $t=0$. One can check that
as $\sigma_++\sigma_-\rightarrow 0$, one gets $\theta \rightarrow
\textrm{arctanh} (g'/f')$, where the prime indicates
differentiation with respect to an arbitrary parameter.

Our strategy consists in the imposition of the two conditions:
\begin{eqnarray}\label{one}
\lim_{\sigma_++\sigma_-\rightarrow
0}x^k=x^{k}_+(\sigma)+x^{k}_-(\sigma) &=& \sum_{n}
a_n^k(0)\exp\big(\imath\frac{2\pi n} {\beta}\textrm{arctanh}\big
(\frac{g'}{f'}\big )(\sigma)\big ),
\end{eqnarray}
\begin{eqnarray}\label{two}
\lim_{\sigma_++\sigma_-\rightarrow
0}\partial_tx^k=\partial_tx^{k}_+(\sigma)+\partial_tx^{k}_-(\sigma)
&=& \sum_n \dot{a}_n^k(0) \exp\big(\imath\frac{2\pi
n}{\beta}\textrm{arctanh}\big (\frac{g'}{f'}\big )(\sigma) \big).
\end{eqnarray}

In this way we get the following simplifications: (i) as we
compare functions on a line we in fact compare functions of a
single variable, (ii) since we choose the line $t=0$, we obtain a
rather simple form on the right-hand side in the form of a
periodic function of $\theta=\arctan(g'/f')$. The only remaining
work to be done is to find the operator $\partial_t$ in the limit
$~\sigma_++\sigma_-\rightarrow 0$.

\subsubsection{Finding $\partial_t$}

Let us start with the general expression for $\partial_t$
\begin{eqnarray}
\nonumber
  \partial_t &=& \frac{\partial_-\theta}{\partial_+t\partial_-\theta-\partial_-t\partial_+
  \theta}~\partial_+ -
  \frac{\partial_+\theta}{\partial_+t\partial_-\theta-\partial_-t\partial_+\theta}
  ~\partial_-~=:~T_-\partial_+
  - T_+\partial_-
\end{eqnarray}
Now, we  evaluate all the components in the above expression:
\begin{equation}
\partial_+\tanh(\theta) = \partial_+\frac{g(\sigma_+)-g(-\sigma_-)}
{f(\sigma_+)-f(-\sigma_-)}\longrightarrow\frac{1}{2}
(\frac{g'}{f'})'~\Big |_{\sigma_+=-\sigma_-}
\end{equation}

\begin{equation}
  \partial_-\tanh(\theta) = \partial_-\frac{g(\sigma_+)-g(-\sigma_-)}
  {f(\sigma_+)-f(-\sigma_-)}\longrightarrow -\frac{1}{2}
  (\frac{g'}{f'})'~\Big |_{\sigma_+=-\sigma_-}
\end{equation}

\begin{eqnarray}
  \partial_+t &=& \partial_+\textrm{sgn}(\sigma_++\sigma_-)
  ~\sqrt{(f(\sigma_+)-f(-\sigma_-))^2-(g(\sigma_+)-g(-\sigma_-))^2}\nonumber\\
  &\longrightarrow &\sqrt{(f')^2-(g')^2}~~~\Big |_{\sigma_+=-\sigma_-}
\end{eqnarray}
\begin{eqnarray}
  \partial_-t &=& \partial_-\textrm{sgn}(\sigma_++\sigma_-)~\sqrt{(f(\sigma_+)
  -f(-\sigma_-))^2-(g(\sigma_+)-g(-\sigma_-))^2}\nonumber\\
  &\longrightarrow &\sqrt{(f')^2-(g')^2}~~~\Big |_{\sigma_+=-\sigma_-}
\end{eqnarray}
One gets
\begin{eqnarray}
  \partial_t=T_-\partial_+
  - T_+\partial_- &\longrightarrow& \frac{1}{2\sqrt{(f')^2-(g')^2}}
  (\partial_++\partial_-)~~\Big |_{\sigma_+=-\sigma_-}
\end{eqnarray}

\subsubsection{Application of the two conditions}

Let us now study the first condition (\ref{one}). At the
singularity we have
\begin{eqnarray}
  x^k_-(\sigma)+x^k_+(\sigma) &=& \sum_n
  a_n^k(0)\exp\big (\imath\frac{2\pi
  n}{\beta}\textrm{arctanh}\big (\frac{g'}{f'}\big )(\sigma)) ,
\end{eqnarray}
which leads to
\begin{eqnarray}
  x^k(\sigma_+,\sigma_-) &=& \sum_n
  a_{n+}^k\exp\big (\imath\frac{2\pi n}{\beta}\textrm{arctanh}\big (\frac{g'}
  {f'}\big )(\sigma_+)\big )\nonumber\\
  &+& \sum_n
  a_{n-}^k\exp\big (\imath\frac{2\pi n}{\beta}\textrm{arctanh}\big (\frac{g'}
  {f'}\big )(-\sigma_-)\big )+F^k
  (\sigma_+)-F^k(-\sigma_-) ,
\end{eqnarray}
where $a_{n+}^k$ and $a_{n-}^k=a_{n}^k(0)-a_{n+}^k$ are constants,
and $F^k$ is a function to be determined.

The second condition (\ref{two}) says that if we act on
(\ref{deter}) with the $\partial_t$ operator and then go to the
limit we will end up with the sum of $\exp\big(\imath\frac{2\pi
n}{\beta}\textrm{arctanh}(\frac{g'}{f'}(\sigma))\big)$ times
constant. It means that
\begin{eqnarray}\label{c1}
  \frac{(\textrm{arctanh}(\frac{g'}{f'}))'}{{\sqrt{(f')^2-(g')^2}}} &=&
  C_1,\\ \label{c2}
\frac{ (F^k)'}{\sqrt{(f')^2-(g')^2}} &=&C_2 ,
\end{eqnarray}
where $C_1\neq 0$ and $C_2$ are constants. From (\ref{c2}) we
conclude that
\begin{eqnarray}
 x^k(\sigma_+,\sigma_-) &=& \sum_n
  a_{n+}^k\exp\big (\imath\frac{2\pi n}{\beta}\textrm{arctanh}\big (\frac{g'}{f'}\big )(\sigma_+)\big )\nonumber\\
  &+&\sum_n
  a_{n-}^k\exp\big (\imath\frac{2\pi
  n}{\beta}\textrm{arctanh}\big (\frac{g'}{f'}\big )(-\sigma_-)\big )\nonumber\\
  &+&c_0^k\textrm{arctanh}\big (\frac{g'}{f'}\big )(\sigma_+)-c_0^k\textrm{arctanh}\big (\frac{g'}{f'}\big
  )(-\sigma_-)
\end{eqnarray}

\subsection{Solutions in the compactified Milne space}

Now let us consider (\ref{c1}). It does not fully determine the
functions $f$ and $g$. This results from the fact that the
topology condition (\ref{sym}), which introduced these functions,
is invariant under the transformation
\begin{equation}
\sigma_+\mapsto\widetilde{\sigma_+}=h(\sigma_+)~,~~~~\sigma_-\mapsto\widetilde{\sigma_-}
=-h(-\sigma_-)
\end{equation}
 where $h$ is  arbitrary. Thus we fix this gauge by introducing a constraint
\begin{equation}
(\textrm{arctanh}(\frac{g'}{f'}))'= \textrm{const}.
\end{equation}
Now from (\ref{c1}) we get
\begin{eqnarray}
 f(\sigma) &=& C\sinh(A\sigma+B)+f_0 ,\\
 g(\sigma) &=& C\cosh(A\sigma+B)+g_0 ,
\end{eqnarray}
where $A$, $B$, $C$, $f_0$ and $g_0$ are arbitrary constants. We
fix the gauge further and put
\begin{equation}
  f(\sigma) = q\sinh(\sigma),~~~~~ g(\sigma) = q\cosh(\sigma)
\end{equation}
where $q$ is an arbitrary constant.

Therefore, the solution reads
\begin{eqnarray}\label{solution}
  x^0 &=& q\sinh(\sigma_+)+q\sinh(\sigma_-),\\
  x^1 &=& q\cosh(\sigma_+)-q\cosh(\sigma_-),\\
  \nonumber\\
  x^k &=& \sum_n a_{n+}^k \exp\big (\imath\frac{2\pi
  n}{\beta}\sigma_+\big )\nonumber\\ &+&\sum_n a_{n-}^k \exp\big (\imath\frac{2\pi
  n}{\beta}\sigma_-\big
  )+{c_0^k}(\sigma_++\sigma_-),
\end{eqnarray}
where $k>1$.

These solutions should satisfy the gauge conditions
(\ref{gauge2}), which in the case of the CM space read
\begin{equation}\label{gauge}
  \partial_+x_k\partial_+x^k = q^2 =
  \partial_-x_k\partial_-x^k .
\end{equation}

At this stage, we can determine  $\theta$ and $t$, from
(\ref{fun1}) and (\ref{fun2}), as functions of $\sigma_+$ and
$\sigma_-$
\begin{eqnarray}
  \theta &=&
  \textrm{arctanh}\Big (\frac{x^1}{x^0}\Big )=\frac{1}{2}\ln\Big (\frac{1+\frac{x^1}{x^0}}{1-\frac{x^1}{x^0}}\Big )=
  \frac{1}{2}\ln\Big (\frac{e^{\sigma_+}-e^{-\sigma_-}}{-e^{-\sigma_+}+e^{\sigma_-}}\Big )\nonumber\\&=&\frac{1}{2}\ln(e^{2\sigma})=\sigma\\
  \nonumber\\
  \frac{t^2}{q^2} &=&
  \big (\sinh(\sigma_+)+\sinh(\sigma_-)\big )^2-\big (\tanh(\sigma_+)-\tanh(\sigma_-)\big )^2\nonumber\\
  &=&-2+2\big (\sinh(\sigma_+)\sinh(\sigma_-)+\tanh(\sigma_+)\tanh(\sigma_-)\big )\nonumber\\=
  &-&2+\cosh(2\tau)=-2+2\big (\sinh(\tau)^2+\cosh(\tau)^2\big )=4\sinh(\tau)^2
\end{eqnarray}
From the last equation we infer that
$\tau=\textrm{arcsinh}(\frac{t}{2q})$. Finally, the solutions as
functions of $t$ and $\theta$ have the form
\begin{eqnarray}\label{gen}
  x^k(t,\theta) &=& \sum_n \Big ( a_{n+}^k e^{\imath\frac{2\pi
  n}{\beta}\textrm{arcsinh}\big (\frac{t}{2q}\big )}+a_{n-}^k e^{-\imath\frac{2\pi
  n}{\beta}\textrm{arcsinh}\big (\frac{t}{2q}\big )}\Big )  \exp\big (\imath\frac{2\pi
  n}{\beta}\theta\big )
  \nonumber\\&+&2c_0^k\textrm{arcsinh}\Big (\frac{t}{2q}\Big ),
\end{eqnarray}
where $n$ denotes $n$-th excitation. The number of arbitrary
constants  in (\ref{gen}) can be reduced by the imposition of the
gauge condition (\ref{gauge}).

Equation (\ref{gen}) defines the solution  corresponding to the
compactification of one space dimension to $S^1$. The solution
corresponding to the compactification to a {\it segment}, can be
obtained from (\ref{gen}) by the imposition of the condition
$x^k(t,\theta) = x^k(t,-\theta)$, which leads to $a^k_{n-} =
a^k_{(-n)-}$ and $a^k_{n+} = a^k_{(-n)+}$, where $\theta \in
[0,\beta/2]$.

The general solution (\ref{gen}) shows that the propagation of a
string through the cosmological singularity is not only continuous
and unique, but also analytic. Solution in the CM space is as
regular as in the case of the Minkowski space.

The imposition of the gauge constraint (\ref{gauge}) on the
infinite set of functions given by (\ref{gen}) produces an
infinite variety of physical states. This procedure goes exactly
in the same way as for a closed string in Minkowski spacetime, but
with a smaller number of degrees of freedom due to the condition
that the string is winding around the compact dimension.

\subsection{Examples of solutions in the CM space}

\subsubsection{The zero-mode state solution}

Let us consider the propagation of a uniformly winding string,
i.e. without $\theta$-dependance, in $d+1$ dimensional spacetime
\begin{eqnarray}
  x^k(t) &=& d_0^k+c_0^k\textrm{arcsinh}\Big (\frac{t}{2q}\Big ) ,\\
  q^2 &=& \frac{1}{4}c_0^kc_{0k} ,
\end{eqnarray}
where the second equation comes from the gauge constraint. Thus we
have
\begin{equation}\label{sol}
  x^k(t) =
  d_0^k+c_0^k\textrm{arcsinh}\Big (\frac{t}{\sqrt{c_0^kc_{0k}}}\Big ).
\end{equation}
Equation (\ref{sol}) coincides with Eq. (18) of our paper
\cite{Malkiewicz:2006bw} describing the propagation of a string in
the zero-mode state, i.e. the lowest energy state.

The velocity of the string, which is calculated with respect to
the `cosmological time' $t$, is given by the formula
$v=(1+\frac{t^2}{c_0^kc_{0k}})^{-1/2}$, where $k>1$. At the
singularity such a string moves with the speed of light.

\subsubsection{The solution with non-zero modes}

In what follows we present the state with one oscillation mode.
The embedding space is five dimensional:
\begin{eqnarray}
x^2(t) &=& d_0^2+c_0^2\textrm{arcsinh}\Big (\frac{t}{2q}\Big )\nonumber \\
x^3(t,\theta) &=& d_0^3+ a \sin\big (\frac{2\pi
  n}{\beta}\big (\theta+\textrm{arcsinh}\Big (\frac{t}{2q}\Big )\big )+\delta\big )+
  a \sin\big (\frac{2\pi
  n}{\beta}\big (\theta-\textrm{arcsinh}\Big (\frac{t}{2q}\Big )\big )-\delta\big )
  \nonumber\\
x^4(t,\theta) &=& d_0^4+ a \cos\big (\frac{2\pi
  n}{\beta}\big (\theta+\textrm{arcsinh}\Big (\frac{t}{2q}\Big )\big )+\delta\big )
  +a \cos\big (\frac{2\pi
  n}{\beta}\big (\theta-\textrm{arcsinh}\Big (\frac{t}{2q}\Big )\big )-\delta\big )
  \nonumber\\
  q^2 &=& \frac{1}{4}(c_0^2)^2+\big (\frac{2a\pi
  n}{\beta}\big )^2 ,
\end{eqnarray}
where the last equation is the gauge constraint. The solution can
be rewritten in a more compact form (after renaming some of the
constants)
\begin{eqnarray}
x^2(t) &=& d_2+c~\textrm{arcsinh}\Big
(\frac{t}{\sqrt{c^2+(\frac{4a\pi
  n}{\beta})^2}}\Big )\nonumber\\
x^3(t,\theta) &=& d_3+ a \sin\big (\frac{2\pi
  n}{\beta}\theta\big )\cos\big (\frac{2\pi
  n}{\beta}\textrm{arcsinh}\Big (\frac{t}{\sqrt{c^2+(\frac{4a\pi
  n}{\beta})^2}}\Big )+\delta\big )\nonumber\\
x^4(t,\theta) &=& d_4+ a \cos\big (\frac{2\pi
  n}{\beta}\theta\big )\cos\big (\frac{2\pi
  n}{\beta}\textrm{arcsinh}\Big (\frac{t}{\sqrt{c^2+(\frac{4a\pi
  n}{\beta})^2}}\Big )+\delta\big )
\end{eqnarray}
Let us calculate the speed of the string
\begin{equation}
v^2:=\partial_tx^k\partial_tx_k=v^2_{\bot}+v^2_{\|}
\end{equation}
and
\begin{eqnarray}
 v^2_{\bot}&=&\frac{c^2}{c^2+(\frac{4a\pi
  n}{\beta})^2+t^2}\\
  v^2_{\|}&=&\frac{(\frac{2a\pi
  n}{\beta})^2\sin^2(\frac{2\pi
  n}{\beta}\textrm{arcsinh}(\frac{t}{\sqrt{c^2+(\frac{4a\pi
  n}{\beta})^2}})+\delta)}{c^2+(\frac{4a\pi
  n}{\beta})^2+t^2}
\end{eqnarray}
where $v_{\bot}$ is the speed of the center of mass and reaches
its maximal value $(1+(\frac{4a\pi n}{\beta c})^2)^{-1/2}$ at the
singularity.  The transverse speed is denoted by  $v_{\|}$ . It
does not depend on $\theta$ since the string is a circle. Its
value is a function of both the cosmological time and the phase of
the oscillation so the total speed $v$  does not necessarily reach
the speed of light at the singularity.  The reason is that the
string, contrary to the previous example,  has some length and
therefore nonzero mass even at the singularity.

\section{Conclusions}

The results presented here show that the classical dynamics of a
test string in a {\it non-zero mode} winding around the
compactified dimension is well defined. There are no ambiguities
in the dynamics near the singularity as is characteristic of
particle dynamics \cite{Malkiewicz:2005ii,Malkiewicz:2006wq}. The
reason is that a string covers the $\theta$-dimension, so the
string does not propagate around this dimension. While a particle
can propagate in this dimension, i.e. a particle has one degree of
freedom connected with the $\theta$-direction; this is not the
case for a string. The property of a string that it is a one
dimensional object makes it possible to go through the singularity
in an unique way. These results are consistent with the analysis
of our previous paper \cite{Malkiewicz:2006bw} dealing with the
string in the {\it zero-mode} state.

We have found the solutions in the CM space with two different
topologies  of the compactified dimension (circle and  segment).
Both solutions are simply related. The latter CM space presents an
interesting  model of two `end of the world' branes, and it has
been  used recently
\cite{Khoury:2001bz,Steinhardt:2001vw,Steinhardt:2001st,Lehners:2008vx}.

The results presented here and in \cite{Malkiewicz:2006bw} open
the door for the examination of the dynamics of extended objects
with dimensionality higher than one. Consideration of the winding
modes is again preferable. This will be shown in our next paper
dealing with the dynamics of a membrane in the compactified Milne
space.

We have not considered transitions between exited  states during
the evolution of the system.  That might give some insight into
the the backreaction problem.  We shall come back to this issue in
next papers.

Our results concern the dynamics of a {\it test} string. By
definition, the test string does not change the background
spacetime. A great challenge is  examination of the dynamics of a
{\it physical} string, i.e. a string that may change its state and
modify the background spacetime during the evolution of the entire
system. Backreaction phenomena cannot be ignored. It has been
shown \cite{Horowitz:2002mw} that, for instance, a single particle
added to a time dependent singular orbifold causes the orbifold to
collapse into a large black hole, which leads finally to the
creation of a big-crunch that would not be followed by a big-bang.
In such a case the compactified Milne space would not make sense
as a model of the neighbourhood of the cosmological singularity in
the context of the cyclic universe scenario. However, the results
of \cite{Horowitz:2002mw} are based mainly on classical general
relativity, which is not suitable for understanding the
microphysics of a black hole. Complete understanding of the
problem requires quantization of the entire system including both
string and embedding spacetime. Work is in progress.

\begin{acknowledgments}
We would like to thank the anonymous referee for finding an error
in the first version of our paper, and for many valuable remarks
and constructive criticism.
\end{acknowledgments}

\end{document}